\documentclass[12pt,aps,
amsmath,
]{article}
\usepackage{bm}
\usepackage{graphicx}

\usepackage{amssymb}

\begin{document}

\title{Infinite Products of Random Isotropically Distributed Matrices}


\author{A.S. Il'yn$^{1}$\footnote{E-mails: asil72@mail.ru, sirota@lpi.ru, zybin@lpi.ru},
V.A. Sirota$^{1}$, K.P. Zybin$^{1,2}$ }
\date
{\small $^1$ P.N.Lebedev Physical Institute of RAS, 119991,
Leninskij pr.53, Moscow, Russian Federation  \\
 $^2$ National Research University Higher
School of Economics, 101000, Myasnitskaya 20, Moscow, Russian
Federation}

\maketitle

\begin{abstract}
Statistical properties of infinite products of random isotropically distributed matrices are investigated.
Both for continuous processes with finite correlation time and discrete  sequences of independent matrices, a formalism that allows to calculate easily the Lyapunov spectrum and generalized  Lyapunov exponents is developed. This problem is of interest to probability theory, statistical characteristics of matrix T-exponentials are also needed for turbulent transport problems, dynamical chaos and other parts of statistical physics.

{\small  Keywords: {Lyapunov exponents \and random matrices \and T-exponential
\and  functional integral \and stochastic equations \and turbulence \and qft} 
 }
\end{abstract}


\section{ Introduction}

Products of random matrices is an important object for many problems of classical and quantum statistical physics \cite{CrisantiPaladinVulpiani}. Mathematical study in this field is mainly devoted to generalization of the law of large numbers, central limit theorem, large deviation theory and other results of the probability theory to the case of non-commutative matrices. The main results in this direction are achieved by Furstenberg, Oseledets, Tutubalin and others \cite{Let1, Oseledets,Let2,Let3}.

The  enlarged law of large numbers states that for a sequence of  products of random independent commutative quantities
\begin{equation}    \label{first}
{q_N} = \prod\limits_{k = 1}^N {\exp \left( {{a_k}} \right)}
\end{equation}
there exists with unitary probability the limit
\[\lambda  = \mathop {\lim }\limits_{N \to \infty } \frac{{\ln {q_N}}}{N} = \left\langle {{a_k}} \right\rangle \]

A generalization of this law to non-commutative random matrices,
\[{Q_N} = \prod\limits_{k = 1}^N {\exp \left( {{A_k}} \right)} \]
involves the notion of Lyapunov spectrum. The main idea is to pick out the diagonal exponentially growing part. It can be done in different ways; for our purposes, the most convenient is the Iwasawa decomposition:
\begin{equation}    \label{Iwasawa}
{Q_N} = {z_N}{\Delta _N}{R_N}
\end{equation}
where $\Delta_N$ is a diagonal matrix with diagonal elements ${\left( {{\delta _N}} \right)_1} \ge 0$,\dots,${\left( {{\delta _N}} \right)_d} \ge 0$, $z_N$ is an upper-triangular matrix with diagonal elements equal to unity,
$R_N$ is an orthogonal matrix,  $R_{ji} R_{jk}=\delta_{ik}$.

Then  with unitary probability the limits
\[{\lambda _s} = \mathop {\lim }\limits_{N \to \infty } \frac{{\ln {{\left( {{\delta _N}} \right)}_s}}}{N},\,s = 1,...,d\]
exist \cite{Let3}. The non-random sequence $\lambda _1 <\dots < \lambda _d$ is called Lyapunov spectrum (LS) of the discrete matrix process $A_k$.
To describe rare processes with exponents significantly different from $\lambda_s$, the theory of large deviations introduces generalized Lyapunov exponents (GLE) defined by
\[{L_s}\left( n \right) = \mathop {\lim }\limits_{N \to \infty } \frac{{\ln \left\langle {{{\left( {\left( {{\delta _N}} \right)_s^{}} \right)}^n}} \right\rangle }}{N}\]
One can see that
\[{\lambda _s} = \frac{{\partial {L_s}\left( 0 \right)}}{{\partial n}}\]

Continuous limit of the random matrix product appears in physical applications as a solution to the linear stochastic equation
\begin{equation} \label{Qequation}
{\partial _t}Q = QA,\,\,Q\left( 0 \right) = \hat 1
\end{equation}
for the evolution matrix. Here $A(t)$ is a stationary $d\times d$ matrix process.%
\footnote{In this paper we restrict ourselves by real $A(t)$ matrices.}
The formal solution to the equation can be written in terms of the
anti-chronological exponential:   
\begin{equation} \label{Q-Texp}
Q\left( t \right) = \mathop T\limits^ +  \exp \left(
{\int\limits_0^t {A\left( \tau \right)d\tau } } \right) =
\sum\limits_n {\frac{1}{{n!}} \int\limits_0^t {d{\tau _1}...d{\tau
_n}\mathop T\limits^ +
 \left( {A\left( {{\tau _1}} \right)...A\left( {{\tau _n}} \right)} \right)} }
\end{equation}
where
\[\mathop T\limits^ +  \left( {A\left( {{\tau _1}} \right) \dots A\left( {{\tau _n}} \right)}
\right) = A\left( {{\tau _{{i_1}}}} \right)...A\left( {{\tau
_{{i_n}}}} \right), \ \quad {\tau _{{i_1}}} \le ... \le {\tau
_{{i_n}}}\] is the antichronological product operator.
 The
alternative way to present the solution is the Volterra
multiplicative integral (\cite{Gantmacher-Volterra}):
\begin{equation} \label{Volterra}
Q\left( t \right) = \prod\limits_{\tau  = 0}^t {\left( {1 + A\left(
\tau \right)d\tau } \right)}
\end{equation}
This allows to treat the solution of  (\ref{Qequation}) as a continuous limit of a random matrix product.
Analogously to the discrete case, with unitary probability there exists the limit
\begin{equation} \label{LS}
{\lambda _s} = \mathop {\lim }\limits_{T \to \infty } \frac{\ln {\delta_s (T)} }T ,\ s = 1,\dots,d
\end{equation}
where $\delta_s (T)= (\Delta (T))_{ss}$ are the elements of the diagonal part of Iwasawa decomposition for the evolution matrix $Q(T)$,
\begin{equation}    \label{Iwasawacont}
{Q(T)} = {z(T)}{\Delta (T)}{R(T)}
\end{equation}
 As in the discrete case, the set $\lambda_1<\dots<\lambda_d$ is called the Lyapunov spectrum (LS) of the continuous process $A(t)$.
 Also generalized Lyapunov exponents are defined by
\begin{equation} \label{GLE}
{L_s}\left( n \right) = \mathop {\lim }\limits_{T \to \infty } \frac{{\ln \left\langle {{{\left( {{\delta_s (T)}} \right)}^n}} \right\rangle }}{T}
\end{equation}
with
\[{\lambda _s} = \frac{{\partial {L_s}\left( 0 \right)}}{{\partial n}}\]

The Lyapunov spectrum was found in \cite{matematicians, matematicians2, GambaKol,Gamba} for the important particular case of isotropic deltacorrelated Gaussian random processes, i.e. processes completely defined by their pair correlation function
$$
\left\langle {{A_{ij}}\left( t \right){A_{kp}}\left( {t'} \right)}
\right\rangle = 2{D_{ijkp}}\delta \left( {t - t'} \right) \ ,
$$
$$
{D_{ijkp}} = a{\delta _{ij}}{\delta _{kp}} + b{\delta _{ik}}{\delta
_{jp}} + c{\delta _{ip}}{\delta _{jk}}
$$
and appeared to be equal to
\begin{equation} \label{lambdaGauss}
\lambda _s^G = D\,\left( {2s - 1 - d} \right)  \  , \qquad D=b+c
\end{equation}

In statistical physics, the LS of continuous processes is mainly needed in problems of turbulence \cite{ZSpre,ufnZS} and turbulent transport  (of temperature, passive density, magnetic field etc.)  \cite{BalkFoux,FalkGawVerg,zeld,koldyn}. In these problems $A(t)$ has the meaning of velocity deformation tensor taken along a particle trajectory in a random velocity field, $A_{ij}(t)= \frac{\partial}{\partial x_i} v_j ({\bf r}(t),t)$.
This random process is known to have non-Gaussian distribution (\cite{nonGauss}). E.g., the third-order velocity increment correlator in turbulence is non-zero (\cite{kolm,Frisch}), which is linked to the energy cascade and to the breaking of time-inversion symmetry. The numerical simulation \cite{GirimajiPope} states that in homogeneous isotropic turbulence the ratio
$\lambda_2/ \lambda_3 $ is $\simeq 1/4$, while for Gaussian processes (\ref{lambdaGauss}) with $d=3$ gives $\lambda_2=0$.
Thus, an approach to calculate the LS and GLE in non-Gaussian processes is needed.

Recently we proposed a functional method that allows to calculate LS and GLE for non-Gaussian isotropic $\delta$-correlated processes \cite{JOSS-2015}. In particular, it was shown that $\lambda_2$ is determined by connected odd-order correlators of $A(t)$.

However, in reality one has finite correlation time. In this paper we present a generalization of the method introduced in \cite{JOSS-2015} for isotropic continuous processes with finite correlation time and isotropic discrete sequences of independent random matrices.

The paper is organized as follows.

In Section 2 we formulate the problem statement and introduce the functional transformation $A(t) \mapsto X(t)$, which allows to pick out the diagonal part of the product. It appears that the LS components coincide with the averages of diagonal elements of $X$.   In Section 3 we discuss the probability measure in $X$ variables.

In Section 4 we introduce formally  the processes with finite correlation time. To calculate the $t\to \infty$ asymptotics of different averages, one can substitute an effective $\delta$-correlated process for the original $A(t)$; this is done  in Section 5.
For isotropic distributions, the substitution  allows to get simple solutions to LS and GLE.

In Section 6 we discuss discrete processes and some details of the passage to continuous limit. We also discuss a way to find LS and GLE of isotropically distributed discrete sequences of independent  matrices.

In the Appendix we  substantiate more formally the limiting process to the effective $\delta$-correlation performed in Section 5.

\section{ $X$-variables  } 

Let $A(t)$ be a stationary random process, $A$ being $d \times d$ real matrices. The statistics of the process is defined by the probability measure in the space of matrix functions, $DA\,P\left[ A \right]$, where
$$
DA \equiv \prod\limits_{t}
{\,\prod\limits_{k,p = 1}^d {d{A_{kp}}\,} ,\,\,\,}
$$
 and $P[A]$  is the probability density functional.
Let $Q(t)$ be a solution of the Eq. (\ref{Qequation}). Our goal is to calculate
 the averages
\begin{equation} \label{averageA}
\left\langle {F\left[ Q \right]} \right\rangle  = \int
{DA\,\,P\left[ A \right]F\left[ Q \right]} \ ,
\end{equation}
 where $F[Q]$ are some functionals which describe the statistics of the $Q$ matrix at large time.
 The formal solution of  (\ref{Qequation}) is given by the antichronological exponential (\ref{Q-Texp}),
 but it is too complicated to use in calculations.  Following \cite{GambaKol,Gamba,JOSS-2015}, we introduce another way to describe the $Q$ matrix.

Recall that $Q(T)$ can be presented in the form of Iwasawa decomposition (\ref{Iwasawacont}). Rewriting (\ref{Qequation})  as
$A = {Q^{ - 1}}{\partial _t}Q$ and substituting (\ref{Iwasawacont}), we obtain
\begin{equation} \label{decAbezX}
A = {R^T}  \left( \rho  + \zeta + \theta \right) R
\end{equation}
where $\rho$ is diagonal, $\zeta$ is upper triangular with zero main diagonal, and $\theta$ is antisymmetric, and the matrices are defined by the equations:
\begin{equation} \label{rho-def}
\rho  = {\Delta^{ - 1}}{\partial _t} \Delta
\end{equation}
\begin{equation} \label{zeta-def}
 \zeta = {\Delta^{ - 1}}{z^{- 1}}\left( {{\partial _t}z} \right) \Delta
\end{equation}
\begin{equation} \label{thetaR}
\theta  = \left( {{\partial _t}R} \right){R^T}
\end{equation}
Averages of the diagonal elements of $\rho$ coincide with the LS (\ref{LS}) of the process $A(t)$ (\cite{Let3}):
 $$
 \lambda_k = \langle \rho_k \rangle \ , \quad \lambda_1< \lambda_2< \dots \lambda_d
 $$

The processes $\rho$,$\zeta$,$\theta $ are asymptotically stationary. Moreover, $z$ stabilizes as $t\to \infty$, which means that
 $z$ tends almost surely to some random limit $z_{\infty}$. To the contrary, $R$ does not stabilize and keeps on walking randomly in SO(d) (\cite{Let3, FalkGawVerg}).
This property of $z$ can be understood by the following consideration: for $\zeta$ from (\ref{zeta-def}) we get
$$
\zeta = \left( \begin{array}{cccc}    0 & \frac{\delta_2}{\delta_1} \left( z^{-1} \partial_t z \right)_{12} &  \dots  &
\frac{\delta_d}{\delta_1} \left( z^{-1} \partial_t z \right)_{1d} \\
0 & 0 & \dots & \frac{\delta_d}{\delta_2} \left( z^{-1} \partial_t z \right)_{2d} \\
0 & 0 & \dots & \dots \\
0 & 0 & 0 & 0
 \end{array} \right)
$$
Since $\lambda_1 <\lambda_2 < \dots \lambda_d$, we expect exponential separation: $\delta_k/ \delta_p \to \infty$ as $t\to \infty$ if $k>p$.
From stationarity $\zeta$, it then follows $\partial_t z \to 0$. This means the stabilization.

\vspace{0.5cm}

 As in \cite{JOSS-2015}, we denote the middle part of (\ref{decAbezX}) by $X(t)$:
 $$
 X = \rho  + \zeta + \theta
 $$
and consider the functional transformation
\begin{equation} \label{decA1}
X \mapsto A = {R^T}[X] \, X \,R[X]  \ ,
\end{equation}
The functional $R$ depends on $X$ non-locally; from (\ref{thetaR}) it follows
\begin{equation} \label{R-Texp}
R\left( t \right) = T\exp \left( {\int\limits_0^t {\theta \left(
\tau  \right)d\tau } } \right)
=
\sum\limits_n {\frac{1}{{n!}} \int\limits_0^t {d{\tau _1} \dots d{\tau
_n}\mathop T
 \left( {\theta \left( {{\tau _1}} \right) \dots \theta \left( {{\tau _n}} \right)} \right)} } \ ,
\end{equation}
\[\mathop T  \left( {\theta\left( {{\tau _1}} \right) \dots \theta \left( {{\tau _n}} \right)}
\right) = \theta \left( {{\tau _{{i_1}}}} \right)...\theta \left( {{\tau
_{{i_n}}}} \right), \ \quad {\tau _{{i_1}}} \ge ... \ge {\tau
_{{i_n}}}\]

The advantage of  the $X$ - variables is, in particular, in dealing with this
simple 'rotational' T-exponent  (\ref{R-Texp}), instead of the
complicated T-exponent (\ref{Q-Texp}).  As we will see below,
in the important case of isotropically distributed $A(t)$ with finite correlation time $t_c<\infty$,
this simpler T-exponent does not contribute to the relations as $T\to \infty$, and the problem of finding the LS can be solved completely.

\section{Statistics of $X$ - variables}

Probability measure of a random process $A(t)$ can be defined by the probability density functional $P[A]$ in the space of matrix functions,
$$
P\left[ A \right] = \exp \left( - S[A] \right)
$$
Here $S[A]$ is the action functional. The transformation (\ref{decA1}) induces the measure in the space of $X$ - variables:
\begin{equation} \label{PX}
P_X\left[ X \right] = N \exp \left( - S\left[ R^T X R \right] \right) J[X]
\end{equation}
where $N$ is the normalizing factor and
\[ J = Det\left( {\frac{\delta {A_{ij}}\left( t \right)}{\delta {X_{kp}}\left( {t'} \right)}} \right)\]
is the Jacobian of the transformation. \cite{JOSS-2015} have shown that it is equal to
\begin{equation} \label{Jac}
J\left[ X \right] = \exp \left( \int   
\mbox{tr} \left(
\eta_0 X (t)  \right) dt  \right) \ ,
\end{equation}
\begin{equation} \label{eta0}
 \left( \eta_0 \right)_{kp} = \,\frac{2k - 1 - d}{2} \delta _{kp}
\end{equation}
(see also \cite{GambaKol,Gamba}).

Generally, use of (\ref{PX}) is rather complicated because it contains the nonlocal functional $R[X]$. However, in the case of isotropic distribution and in the limit $T\to \infty$ the situation may be simplified significantly. Preparatory to proceeding to this important case, we recall several important definitions (see, e.g., \cite{Klyackin}).

\section{Correlation functions of the processes with finite correlation time}

The Fourrier transform of $P[A]$ is called characteristic functional:
\begin{equation}  \label{Z}
\begin{array}{r}   \displaystyle
Z\left[ {\eta \left( t \right)} \right] = \int {DA\,\exp \left( -S[A]+i{ \int 
{tr\left(
 \eta(t) A(t) \right) dt} }  \right)} \\
= \left\langle {\exp \left(
i {\int
{tr\left( {\eta \left( t \right)A\left( t \right)}
\right)dt} }
 \right)} \right\rangle
\end{array}
 \end{equation}
  Here $\eta(t)$ is a $d\times d$-matrix element adjoint to $A(t)$.
$Z$ is the generating functional for the correlation functions:
\begin{equation} \label{AZ}
\langle A_{ij} \left( t_1 \right) \dots A_{kp} \left( t_n \right)
\rangle  = 
\frac{1}{i^n}
\frac{\delta }{\delta \eta _{ij}\left(
t_1 \right)} \dots \frac{\delta }{\delta \eta _{kp}\left( t_n
\right)}Z [{0}]
\end{equation}
From normalization condition $\left \langle 1 \right \rangle =1$ it follows $Z[0]=1$.

It is convenient to introduce the generating functional  of connected (irreducible) correlation functions defined by
\begin{equation} \label{W}
W[\eta] = \ln Z[\eta]
\end{equation}
It is also called cumulant functional. Expanding it into a series, we get
$$
W[\eta] =  \sum \limits_n \frac1{n!} \int W^{(n)}_{ij \dots kp} (t_1, \dots, t_n) \eta_{ij}(t_1) \dots \eta_{kp}(t_n) dt_1 \dots dt_n
$$
Then $n$-point connected correlation functions can be defined by
\begin{equation} \label{AAc}
\left\langle {{A_{ij}}\left( {{t_1}} \right)...{A_{kp}}\left(
{{t_n}} \right)} \right\rangle_c =
\frac 1{i^n}
\frac{\delta }{\delta \eta _{ij} (t_1)}  \dots \frac{\delta }{\delta \eta _{kp} (t_n)} W[0] =
\frac 1{i^n} W^{(n)}_{ij \dots kp} (t_1, \dots, t_n)
\end{equation}

Hereafter we consider homogeneous processes with finite correlation time. This means that connected correlation functions depend only on differences of times $|t_1-t_k|$ and decrease quickly as $|t_1-t_k|>t_c$. So,
$$
W^{(n)}_{ij \dots kp} (t_1, \dots, t_n) = W^{(n)}_{ij \dots kp} (t_1-t_2, \dots, t_1-t_n) \to 0 \ , \qquad |t_1-t_k|>t_c
$$

Since we are interested in asymptotic behavior of the averages (\ref{averageA}), we only need to calculate them for $T \gg t_c$.
To this purpose, we can substitute an effective $\delta$-process for $A(t)$. In the case of isotropic processes, i.e. the processes statistically invariant under global rotations,
\begin{equation} \label{isotrop}
P \left[ {{O^T} A O} \right] = P\left[ A \right] \ \  \forall O \in SO(d) \ ,
\end{equation}
this allows to avoid non-locality and to solve the problem completely.
\footnote{In non-isotropic case, the effective $\delta$-process does not help much because $X(t)$ remains non-local even for delta-correlated $A(t)$ (see details in the Appendix). }

In the next Section, we introduce the effective $\delta$-process and make use of it; the details of the substitution and its correctness are discussed in the Appendix.

\section{Effective $\delta$-process}

The cumulant functional of the effective process can be obtained from the connected correlation functions of the original process
 by the transformation:
$$
W^{(n)}_{ij \dots kp} (t_1-t_2, \dots, t_1-t_n) \longrightarrow  {W^{\it eff}}^{(n)}_{ij \dots kp} =
{w}^{(n)}_{ij \dots kp} \delta(t_1-t_2) \dots \delta(t_1-t_n) \ ,
$$
\begin{equation} \label{cumul1}
{w}^{(n)}_{ij \dots kp} = \int W^{(n)}_{ij \dots kp} (t_1-t_2, \dots, t_1-t_n) dt_2 \dots dt_n
\end{equation}
The effective cumulant functional $W^{\it eff}[\eta]$ is  then given by
\begin{equation} \label{Wefflocal}
W^{\it eff} [\eta] = \int 
{w} (\eta(t)) dt
\end{equation}
where
\begin{equation} \label{cumul2}
{w} (\eta)= \sum \frac 1{n!} {w}^{(n)}_{ij \dots kp} \eta^n
\end{equation}
is called cumulant function.
Thus, the connected correlation functions of the effective process are
\begin{equation} \label{AAc-eff}
\left\langle {{A^{\it eff}_{ij}}\left( {{t_1}} \right) \dots {A^{\it eff}_{kp}}\left(
{{t_n}} \right)} \right\rangle_c =
\frac 1{i^n}
\frac{\partial }{\partial \eta _{ij} }  \dots \frac{\partial }{\partial \eta _{kp} } {w}(0) \, \delta(t_1-t_2) \dots \delta(t_1-t_n)
\end{equation}
The action is related to the cumulant functional via the Fourrier transform
(\ref{Z}) and (\ref{W}).
Hence, from (\ref{Wefflocal}) it follows that the effective action can be written as an integral of a local Lagrangian:
\begin{equation} \label{localaction}
S^{\it eff} \left[ A^{\it eff} (t) \right] =  \int 
{L} \left( A^{\it eff}(t) \right) dt
\end{equation}
So, the probability density of the effective process takes a form of a continuous product of local probability densities, which corresponds to the independence of the values of $A^{\it eff}$ taken at different time moments.

For example, for the Gaussian $\delta$-correlated process one has
\begin{equation} \label{wGauss}
w^G(\eta) = - \eta_{ij} D_{ijkp} \eta_{kp}  \  ,
\end{equation}
$$
\qquad  L^G(A) = \frac 14 A^{\it eff}_{ij} D^{-1}_{ijkp} A^{\it eff}_{kp}
$$
where $D_{ijkp} $ is $d^2\times d^2$ matrix; according to (\ref{AAc-eff}),
$$
2 D_{ijkp} \delta (t-t') =
\left\langle {{A^{\it eff}_{ij}}\left( {{t}} \right) {A^{\it eff}_{kp}}\left(
{{t'}} \right)} \right\rangle_c   =
\left\langle {{A^{\it eff}_{ij}}\left( {{t}} \right) {A^{\it eff}_{kp}}\left(
{{t'}} \right)} \right\rangle
$$

\vspace{0.5cm}
We note that cumulant function and Lagrangian are two alternative ways to determine a $\delta$ -correlated random process. However, a typical problem of local  quantum field theory with interaction is that for non-gaussian processes, finite Lagrangian produces ultraviolet divergences in correlation functions. And vice versa, 'good' correlation functions imply well-defined cumulant function and divergent Lagrangian. Since correlation functions is the only quantity that has physical meaning, it is more convenient to describe $\delta$-processes by a cumulant function, thereby to avoid divergency problems. One more way to work with delta-correlated non-gaussian processes without divergences is introducing of smoothed variables, as it was done by \cite{JOSS-2015}.

\vspace{0.7cm}
Now we proceed to isotropic $\delta$-processes (\ref{isotrop}). For these processes, Lagrangian and cumulant functions are both rotationally invariant:
$$
{L} \left( {{O^T} A^{\it eff} O} \right) = 
{L} \left( A^{\it eff} \right) \ \mbox{and} \ w \left( {{O^T} \eta O} \right) = w\left( \eta \right) \ \  \forall O \in SO(d) \ ,
$$
Thus, they contain only invariant combinations of variables:
\begin{equation} \label{isot}
\begin{array}{l}   \displaystyle
{L} ( A^{\it eff} ) = 
{L}\left( {tr\,A^{\it eff}, \, \, tr\left({A^{\it eff}} \right)^2,\,\,tr \left( A^{\it eff} {{A^{\it eff}}}^T \right),\,tr\left( A^{\it eff}\right) ^3, \dots} \right) \ , \\ 
w ( \eta ) = w\left( {tr\,\eta,\,\,tr\,{\eta^2},\,\,tr\,\eta{\eta^T},\,tr\,{\eta^3}, \dots} \right)
\end{array}
\end{equation}
From (\ref{PX}) and (\ref{localaction}) it then follows that the non-local variable $R[X]$ is omitted from (\ref{PX}), and
hence the  Lagrangian $
{L}_X$ and the cumulant function $w_X$ of $X^{\it eff}$ -variables are also local, and have also a form as (\ref{isot}). Thus, if $A^{\it eff}$ is isotropic and $\delta$-correlated, the corresponding $X^{\it eff}$ is also isotropic and $\delta$-correlated.

Thus, from (\ref{PX}) we get
\begin{equation} \label{PXmod}
P^{\it eff}_X\left[ X^{\it eff} \right] = N \exp \left( - \int L\left(  X^{\it eff}  \right) dt + \int tr (\eta_0 X^{\it eff}) dt  \right)
\end{equation}
Performing the functional Fourrier transform, we get
$$
\begin{array}{l} \displaystyle
Z^{\it eff}_X [\eta(t)] =  Z^{\it eff} [\eta(t) - i \eta_0]  \left({Z^{\it eff}}\right)^{-1} [-i\eta_0] \ , \\
W^{\it eff}_X [\eta(t)] =  W^{\it eff} [\eta(t) - i \eta_0] - W^{\it eff} [-i\eta_0] \
\end{array}
$$
and eventually
$$
w_X (\eta) =  w (\eta - i \eta_0) - w (-i\eta_0)
$$
So, to get correlation functions of $X^{\it eff}$-variables we can use the same generating functionals as for $A^{\it eff}$-variables,
the only difference is the shift of the point:
\begin{equation} \label{Xw}
\left\langle {{X^{\it eff}_{ij}}\left( {{t_1}} \right)...{X^{\it eff}_{kp}}\left(
{{t_n}} \right)} \right\rangle _c =   \frac 1{i^n}\frac{\partial }{\partial
\eta _{ij}} \dots   \frac{\partial }{\partial \eta _{kp}}
w(-i \eta_0)  \delta(t_1-t_2) \dots
\delta(t_1-t_n)
 \end{equation}
In particular, from (\ref{Xw}) we get simple expressions for LS and GLE (\cite{JOSS-2015}):
\begin{equation}  \label{lamb1}
\lambda_s= \left\langle {{X_{ss}}} \right\rangle=
\left\langle {{X^{\it eff}_{ss}}} \right\rangle  = -i  \frac{\partial }{{\partial
{\eta _{ss}}}} w_X \left(  0  \right) =-i  \frac{\partial }{{\partial
{\eta _{ss}}}} w\left( { -i \eta_0 } \right)
\end{equation}
\begin{equation}  \label{lamb2}
L_s (n) = -i   w_X \left(  0, \dots,0, \eta_{ss}=-in, 0, \dots,0  \right)
\end{equation}
For Gaussian cumulant function (\ref{wGauss}), we get the known result (\ref{lambdaGauss}).

To summarize the recipe for   finite time-correlated processes, one first has to find the cumulant function  $w(\eta)$ of
the corresponding $\delta$-correlated process by means of (\ref{cumul1}), (\ref{cumul2}), and then substitute it into
(\ref{lamb1}),(\ref{lamb2}).  

 General properties of LS and GLE of
isotropic $\delta$-processes were described in details in \cite{JOSS-2015}. Here we only focus on one property important for the next
Section: for isotropic $\delta$-correlated processes the functions $\left \langle \ln \delta^{\it eff}_s (T) \right \rangle /T$ and
$ \ln \left \langle \left( \delta^{\it eff}_s (T) \right) ^n \right \rangle /T $ do not depend on $T$ (\cite{JOSS-2015}). \label{pol}
\footnote{This is a special property of the Iwasawa decomposition; for, e.g., polar decomposition of the evolution matrix,
$Q=rDR$, $r,R \in SO(d)$, $D=diag\{ D_1, \dots, D_d \}$ this is not so, the functions
$\left \langle \ln D_s (T) \right \rangle /T$ and $ \ln \left \langle \left( D_s (T) \right) ^n \right \rangle /T $ depend on $T$
even in $\delta$-correlated $A$ - processes, and asymptotically tend to $\lambda_s$ and $L_s(n)$ as $T \gg (\lambda_d-\lambda_1)^{-1}$
(see \cite{FalkGawVerg}). }
 Therefore,
for these processes one can omit the sign of limit in (\ref{GLE}). To the contrary, for the processes with finite correlation time
these quantities depend on $T$, and only their limits as $T \to \infty$  (or, more accurately, $T \gg t_c$) coincide with the LS
and GLE of the effective $\delta$-process.

\section{Discrete processes}

 We have analyzed continuous products of matrices by means the Eq.(\ref{Qequation}), which they have been shown to satisfy.
On the other hand, continuous process can be treated as a limit of a discrete process $A(t_k)$ with a step $\Delta t=t_{k+1}-t_k$.
Is it possible to use the same method for discrete products?  To this effect, the discrete product should satisfy a difference
equation close to the differential equation (\ref{Qequation}). In particular, the matrices  $A(t_k)$ and $A(t_{k+1})$ must not
differ too much. More formally, the limit $\Delta t \to 0$ must be taken at finite correlation time $t_c$.

So, the same method is applicable directly to discrete matrix products satisfying the condition $\Delta t \ll t_c \ll T$.
The classical problem  concerning the product of independent matrices  does not
satisfy this requirement, since in the case $\Delta t=1$ and $t_c=0$, and the consecutive matrices may differ strongly.
However, the results achieved for continuous products can still help in solving this problem.

Let $A_1 \dots A_N$ be a sequence of  random matrices with the same isotropic probability density $p(A_k)$.  Consider the product
\begin{equation}  \label{Qdiscrete}
Q_N = exp \left( A_1 \right) \dots exp \left( A_N \right)
\end{equation}
Again we are interested in the LS and GLE:
\[{\lambda _s} = \mathop {\lim }\limits_{N \to \infty } \frac{{\ln {{\left( {{\delta _N}} \right)}_s}}}{N} \ ,\quad
{L_s}\left( n \right) = \mathop {\lim }\limits_{N \to \infty } \frac{{\ln \left\langle {{{ {\left( {{\delta _N}}
\right)_s} }^n}} \right\rangle }}{N} \ , \quad s = 1,...,d\]
where, as before, $(\delta_N)_s$ are the elements of the diagonal matrix $\Delta_N$ from the Iwasawa decomposition of $Q_N$
(\ref{Iwasawa}).

Let $B(t)$ be a stationary $\delta$-correlated isotropic random process such that probability density of
$T \exp  \int \limits_{k-1}^k B(t) dt $ coincides with the probability density of $\exp \left( A_k \right) $. In one-dimensional case ($d=1$), its existence is easily proved by writing the explicit form of the cumulant function:
 $w(\eta)=\ln \left \langle \exp (\left( i \eta A \right) \right \rangle = \ln \int dA \, p(A) \exp \left( i \eta A \right)$.
 For bigger dimensions, the existence of $B$ for arbitrary process requires additional arguments.

Roughly speaking, one can say that the process $B(t)$ presents a 'continuous model' of the discrete process $A_k$. The problem of
counting LS and GLE then reduces to the corresponding problem for the continuous process,
\begin{equation}  \label{B1}
 Q_N = T \exp  \int \limits_{0}^N B(t) dt
\end{equation}
 On the other hand, as we noticed in the end of the previous Section,  the values
 $\left \langle \ln \delta_N \right \rangle /N$ and
$ \ln \left \langle \left( \delta_N \right) ^n \right \rangle /N $ do not depend on $N$.
In this sense, isotropically distributed matrices behave like commutative quantities (\ref{first}).

Thus, to calculate LS and GLE,
one can put $N=1$: 
\begin{equation}  \label{discret-result}
\lambda_s= \left \langle \ln \left( \delta_1 \right)_s \right \rangle   \ , \quad
L_s (n) =  \ln \left \langle { \left( \delta_1 \right)_s }  ^n \right \rangle
\end{equation}
We note that the explicit form of the auxiliary  process $B(t)$ does not contribute to the result. Since Iwasawa decomposition for
the matrix $Q_1=\exp (A_1)$, and in particular $\Delta_1 (A_1)$, is univocal, it is enough to know $p(A_1)$
or probability density of $Q_1$ to calculate the averages (\ref{discret-result}).

As a simple illustration, we now use (\ref{discret-result}) to calculate LS and GLE  for a sequence of isotropically distributed
 symmetric traceless $2 \times 2$ matrices.  They can be presented in the form
 $$
A_k = \left(  \begin{array}{rr} \cos \phi_k & - \sin \phi_k \\ \sin \phi_k &  \cos \phi_k  \end{array} \right)^T
 \left(  \begin{array}{cc} -a_k & 0 \\  0 & a_k  \end{array} \right)
\left(  \begin{array}{rr} \cos \phi_k & - \sin \phi_k \\ \sin \phi_k &  \cos \phi_k  \end{array} \right) \ ,
 $$
From isotropy it follows that the probability density $p(a_k, \phi_k)=p(a_k)$ does not depend on $\phi_k$.
Then
\begin{equation}  \label{B-Q}
Q_1=\exp (A_1) = \left(  \begin{array}{rr} \cos \phi_1 & - \sin \phi_1 \\ \sin \phi_1 &  \cos \phi_1  \end{array} \right)^T
 \left(  \begin{array}{lc} e^{-a_1} & 0 \\  0 & e^{a_1}  \end{array} \right)
\left(  \begin{array}{rr} \cos \phi_1 & - \sin \phi_1 \\ \sin \phi_1 &  \cos \phi_1  \end{array} \right) \ ,
\end{equation}
The Iwasawa decomposition of $Q_1$ gives
$$
\left( \delta_1 \right)_s (a_1, \phi_1) = \left(  \mbox{ch} (2a_1) + \cos (2 \phi_1) \mbox{sh} (2 a_1) \right) ^{(-1)^s /2}
$$
Then from (\ref{discret-result}) we get
$$
\lambda_s =  \frac 1{2\pi} \int \limits_{-\infty}^{\infty} d a_1 p(a_1) \int \limits _0^{2\pi} d \phi_1 \ln \left( \delta_1 \right)_s
= (-1)^s \int \limits_{-\infty}^{\infty} d a_1 p(a_1) \ln \left( \mbox{ch} \, a_1 \right)
$$
One can see that if the integral accumulates at large $a_1$, i.e. if $ \langle \left| a_1 \right| \rangle \gg 1$ then $\lambda_s \simeq (-1)^s \langle |a_1| \rangle $.
This fact has a natural explanation. Actually,
$\lambda_s$ can be derived in
terms of polar decomposition of $Q_N$ as well:
$\lambda_s = \lim \limits_{N\to \infty}\left \langle \ln  (D_N)_s \right \rangle /N$
where $D_N$ is the diagonal matrix.
According to the footnote on page \pageref{pol}, the limit is saturated as soon as $N \gg |\lambda_2-\lambda_1|^{-1}$; hence,
for large values of $|\lambda|$  this condition is already satisfied for $N=1$.
  On the other hand,
(\ref{B-Q}) represents the polar decomposition of $Q_1$, and $D_{1,2}=\exp\left( \mp |a_1|\right)$, so
$\lambda_s \simeq  \left \langle \ln  (D_1 ) _s  \right \rangle =
 \left \langle  (-1)^s  \left| a_1  \right|  \right \rangle$.

\section{Conclusion}

In the paper we have considered the Lyapunov spectrum and general Lyapunov exponents of a continual product of isotropically
distributed finite time-correlated random  matrices and of a discrete product of independent (and isotropically distributed)
matrices. We have shown that in the case of finite time correlation the LS and GLE coincide with those of some effective
delta-correlated process. Simple relations  (\ref{cumul1}), (\ref{cumul2}), (\ref{lamb1}) and (\ref{lamb2}) have been
obtained to derive LS and GLE  and other averages
in terms of connected correlation
functions of  the original process.  For discrete products we have also obtained a simple expression (\ref{discret-result})
relating LS and GLE to the
probability density of the multiplied matrices.

This work is supported by the RAS program 'Nonlinear dynamics in
mathematical and physical sciences'.

\vspace{1cm}

\section{Appendix.  Effective $\delta$-process}

In the Appendix, we show that the approximation of $\delta$-correlated process is applicable whenever
we are interested in the evolution of a process at the time much bigger than the correlation time.
Local Lagrangian is badly defined for non-Gaussian processes, so we will use the formalism of cumulant functionals.

For simplicity, we first consider a one-dimensional stationary isotropic random process $A(t)$ with finite correlation time $t_c$.
In accordance with Section 4, its statistics is defined by the cumulant functional:
 \begin{equation}  \label{Wap}
 W\left[ {\eta \left( t \right)} \right] =
\ln \left\langle {\exp \left(
 i {\int
 {\left( {\eta \left( t \right)A\left( t \right)}
 \right)dt} }
  \right)} \right\rangle
  \end{equation}
  with normalization condition  $ W[0] =0  $.
  It can be expanded into a Taylor series
\begin{equation} \label{Apl3}
W[\eta] =  \sum \limits_n \frac1{n!} \int W^{(n)} (t_1-t_2, \dots, t_1-t_n) \eta(t_1) \dots \eta(t_n)
 dt_1 \dots  dt_n
\end{equation}
where the coefficients are related to the connected correlation functions of $A$ according to
$$ 
 W^{(n)} (t_1, \dots, t_n)  =  {i^n}
\left\langle {{A}\left( {{t_1}} \right)...{A}\left( {{t_n}} \right)} \right\rangle_c
$$ 
Since $t_c$ is the correlation time, the functions $ W^{(n)} (t_1, \dots, t_n) $ must decrease rapidly as
$|t_1-t_k|\gg t_c$. We are interested in time scales $t\sim T\gg t_c$. Thus, we need only rather 'slow' functions $\eta(t)$ in
(\ref{Wap}). To single them out, we change the variables in (\ref{Apl3}):
$$ \tau = t/ T  \ , \quad \mu(\tau) = \eta (\tau T)  $$
(The corresponding change of $A$ is $a(\tau)=T A(\tau T)$, this is just a rescaling of time and the dimensional variable $A$.)

Then as $t\to \infty$, we have
\[\begin{array}{l}
W\left[ {\eta \left( t \right)} \right] =  W\left[ {\mu \left[ \tau  \right]} \right] \\
 = \sum\limits_{n = 1}^{\infty } {\frac{1}{{n!}}\int\limits_{ - \infty }^\infty  {d{\tau _1}...d{\tau _n}\,\,{T^n}{W^{(n)}}\left( {T\left( {{\tau _1} - {\tau _2}} \right),...,T\left( {{\tau _1} - {\tau _n}} \right)} \right)\mu \left( {{\tau _1}} \right)...\mu \left( {{\tau _n}} \right) \to } } \\
 \to T\sum\limits_n {\int\limits_{ - \infty }^\infty  {d{\tau _1}...d{\tau _n}\,\frac{{{w^{(n)}}}}{{n!}}\,\delta \left( {{\tau _2} - {\tau _1}} \right)} ...\delta \left( {{\tau _n} - {\tau _1}} \right)\mu \left( {{\tau _1}} \right)...\mu \left( {{\tau _n}} \right)}
\end{array}\]
where
 \begin{equation}  \label{Ap4}
{w^{(n)}} \equiv \int\limits_{ - \infty }^\infty  {d{t_2}...d{t_n}\,\,{W^{(n)}}\left( {{t_1} - {t_2},...,{t_1} - {t_n}} \right)}
\end{equation}

Thus, the statistics at large time scales is described by the effective $\delta$-process with the cumulant functional
 \begin{equation}  \label{Ap5}
{W^{\it eff}}\left[ {\mu \left[ \tau  \right]} \right] = T\,\sum\limits_n {\frac{{{w^{(n)}}}}{{n!}}\int\limits_{ - \infty }^\infty  {{{\left( {\mu \left( \tau  \right)} \right)}^n}d\tau }  \equiv } T\int\limits_{ - \infty }^\infty  {w\left( {\mu \left( \tau  \right)} \right)d\tau }  
 \end{equation}
and connected correlation functions
\[W_{\it eff}^{\left( n \right)}\left( {{\tau _2} - {\tau _1},\dots,{\tau _n} - {\tau _1}} \right) = T{w^{\left( n \right)}}\delta \left( {{\tau _2} - {\tau _1}} \right) \dots \delta \left( {{\tau _n} - {\tau _1}} \right)\]  
The function $w(\mu)$ is just the cumulant function defined in (\ref{cumul2}).

Thus, in one-dimensional case the procedure of substitution the effective cumulant functional for the initial one to calculate
long-time asymptotics is correct.

For a simple example, we consider the equation
 \begin{equation}  \label{Ap7}
Q\left( T \right) = \exp \left( {\int\limits_0^T {A(t)dt} } \right) 
 \end{equation}
 and fine the $T\gg t_c$ asymptotics of the moments $ \left\langle {{Q^\alpha }(T)} \right\rangle $.
 According to (\ref{Wap}), (\ref{Ap7}),
 \[\left\langle {{Q^\alpha }(T)} \right\rangle  = \exp W\left[ { - i\alpha \Theta \left( t \right)\Theta \left( {T - t} \right)}
 \right]\]
 where $\Theta$ is the Heaviside function.
  Hence, as $T\to \infty$ we have
\[\left\langle {{Q^\alpha }(T)} \right\rangle  = \exp \left( {{W_{\it eff}}\left[ { - i\alpha \Theta \left( \tau  \right)\Theta
\left( {1 - \tau } \right)} \right]} \right)\]
   From (\ref{Ap5}) and (\ref{Ap7}), taking into account the normalization $w(0)=0$, we get
\[\left\langle {{Q^\alpha }(T)} \right\rangle  = \exp \left( {w\left( { - i\alpha } \right)T} \right) \  \ as \  T\to \infty \] 

We now proceed to $d\times d$ matrix processes. Again, in accordance with (\ref{Z}), (\ref{W}) we define the statistics of $A$ by
means of the cumulant functional:
\[W\left[ \eta  \right] = \ln \left\langle {\exp \left( {i\int {tr\left( {\eta A} \right)dt} } \right)} \right\rangle \] 
 The probability density is then equal to
\[P\left[ A \right] = {N'}\int {D\eta \,\exp \left( {W\left[ \eta  \right] - i\int\limits_{}^{} {tr\left( {\eta A} \right)dt} } \right)} \ , \] 
where $N'$ is a normalization constant.

The change of functional  variables $A$ to $X$ (\ref{decA1}) generates the probability density in $X$ variables (see (\ref{PX}),
(\ref{Jac})):
\[{P_X}\left[ X \right] =  {N} {P_A}\left[ {{R^T}\left[ X \right]X\,R\left[ X \right]} \right]\exp \int {dt\,tr\left( {{\eta _0}X} \right)} \]
so we get
\[
\begin{array}{r}
{W_X}\left[ \eta  \right] = \ln \int DX\,D\eta '\,\exp \left( W\left[ {\eta '} \right] - i\int\limits_{ - \infty }^{\infty}  {tr\left( \eta '{R^T}\left[ X \right]X\,R\left[ X \right] \right)dt} \right. \\
 \left. + i\int\limits_{ - \infty }^\infty  {tr\left( {\left( {\eta  - i{\eta _0}} \right)X} \right)dt}  \right) - C
  \end{array} \]
the constant $C$ providing $W[0]=0$.

In the inner integrel ($X$ is fixed) we replace $\eta'$  with
$ {R^T}\left[ X \right]\eta 'R\left[ X \right]$. This is an orthogonal transformation, thus
\[
\begin{array}{r}
{W_X}\left[ \eta  \right] = \ln \int DX\,D\eta '\,\exp \left( {W_A}\left[ {{R^T}\left[ X \right]\eta 'R\left[ X \right]} \right] - i\int\limits_{ - \infty }^\infty  {tr\left( {\eta 'X} \right)dt} \right. \\
\left. + i\int\limits_{ - \infty }^\infty  {tr\left( {\left( {\eta  - i{\eta _0}} \right)X} \right)dt}  \right) - C
\end{array} \] 

As in the one-dimensional case, we are interested in the case $t\sim T \gg t_c$. So we rescale the variables:
$$ \tau = t/ T  \ , \quad \mu(\tau) = \eta (\tau T)  \ , \quad  x(\tau)= T \, X(\tau T) $$
Then
\[R\left[ {X(t),t} \right] = T\exp \int\limits_0^t {dt'\theta \left( {t'} \right)}  = R\left[ {x(\tau ),\tau } \right]\]
and
\[
\begin{array}{r}
{W^{\it eff}_{X}}\left[ {\mu \left( \tau  \right)} \right] = \ln \int DxD\mu '\,\exp \left( {W^{\it eff}_{A}}\left[ {{R^T}\left[ {x,\tau } \right]\mu '\left( \tau  \right)R\left[ {x,\tau } \right]} \right] \right. \\
\left. + i\int\limits_{ - \infty }^\infty  {tr\left( {\left( {\mu \left( \tau  \right) - \mu '\left( \tau  \right) - i{\eta _0}} \right)x\left( \tau  \right)} \right)d\tau }  \right) - C'
\end{array}  \]
where $W^{\it eff}$ is defined in the same way as in the one-dimensional case.
The isotropy condition implies
\[{W^{\it eff}_{A}}\left[ {{R^T}\left[ {x,\tau } \right]\mu '\left( \tau  \right)R\left[ {x,\tau } \right]} \right] = {W^{\it eff}_{A}}\left[ {\mu '\left( \tau  \right)} \right]\]
thus
\[
\begin{array}{rcl}
{W^{\it eff}_{X}}\left[ {\mu \left( \tau  \right)} \right] &=& \ln \int DxD\mu '\,\exp \left( W^{\it eff}_A\left[ \mu '\left( \tau  \right) \right] \right. \\
 &+& \left. i\int\limits_{ - \infty }^{\infty}  {tr\left( {\left( \mu \left( \tau  \right) - \mu '\left( \tau  \right) - i \eta _0 \right)x\left( \tau  \right)} \right)d\tau }  \right) - C'
\end{array} \]
The integral over $x$ gives the $\delta$-function
$\prod\limits_\tau  {\delta \left( {\mu \left( \tau  \right) - \mu '\left( \tau  \right) - i{\eta _0}} \right)} $.
Integrating over $\mu'$, we eventually get
\[{W^{\it eff}_{X}}\left[ {\mu \left( \tau  \right)} \right] = {W^{\it eff}_{A}}\left[ {\mu \left( \tau  \right) - i{\eta _0}} \right] - {W^{\it eff}_{A}}\left[ { - i{\eta _0}} \right]\] 

We see that in the isotropic case, effective $\delta$-correlated cumulant functionals of $A$ and $X$ variables
differ only by the  shift $-i\eta_0$ of the argument.

\end{document}